\documentclass[useAMS,usenatbib]{mn2e}
\usepackage{graphicx,amsmath,multirow,amssymb}
\usepackage{subfig}
\usepackage{natbib}
\newcommand{\comment}[1]{}

\def\simgt{\lower.5ex\hbox{$\; \buildrel > \over \sim \;$}}
\def\simlt{\lower.5ex\hbox{$\; \buildrel < \over \sim \;$}}

\title[Massive AGBs in the LMC]{The Large Magellanic Cloud as a laboratory for Hot Bottom Burning in
massive Asymptotic Giant Branch stars}
\author[Ventura et al.]{P. Ventura$^1$, A. I. Karakas$^{2}$, F. Dell'Agli$^{1,3}$, M.~L. Boyer$^{4,5}$,
\newauthor
D. A. Garc\'{\i}a--Hern\'andez$^{6,7}$, M. Di Criscienzo$^1$, R. Schneider$^1$  \\
$^1$INAF -- Osservatorio Astronomico di Roma, Via Frascati 33, 00040, Monte Porzio Catone (RM), Italy \\
$^2$Research School of Astronomy and Astrophysics, Australian National University, Canberra, ACT 2611, Australia \\
$^3$Dipartimento di Fisica, Universit\`a di Roma ``La Sapienza'', P.le Aldo Moro 5, 00143, Roma, Italy \\
$^4$Observational Cosmology Lab, Code 665, NASA Goddard Space Flight Center, Greenbelt, MD 20771, USA \\
$^5$Oak Ridge Associated Universities (ORAU), Oak Ridge, TN 37831, USA  \\
$^{6}$Instituto de Astrof\'{\i}sica de Canarias, E-38200 La Laguna, Tenerife, Spain \\
$^{7}$Departamento de Astrof\'{\i}sica, Universidad de La Laguna (ULL), E-38206 La Laguna, Tenerife, Spain\\
}

\begin{document}

\date{Accepted, Received; in original form }

\pagerange{\pageref{firstpage}--\pageref{lastpage}} \pubyear{2012}

\maketitle

\label{firstpage}

\begin{abstract}
We use Spitzer observations of the rich population of Asymptotic Giant Branch stars in 
the Large Magellanic Cloud (LMC) to test models describing the internal structure and 
nucleosynthesis of the most 
massive of these stars, i.e. those with initial mass above $\sim 4M_{\odot}$.
To this aim, we compare Spitzer observations of LMC stars with the theoretical tracks of
Asymptotic Giant Branch models, calculated with two of the most popular  
evolution codes, that are known to differ in particular for the treatment of convection.

Although the physical evolution of the two models are significantly different,  
the properties of dust formed in their winds are surprisingly similar, as is their position
in the colour--colour (CCD) and colour--magnitude (CMD) diagrams obtained with the 
Spitzer bands. This model independent result allows us to select a well
defined region in the ($[3.6]-[4.5], [5.8]-[8.0]$) plane, populated by AGB stars experiencing
Hot Bottom Burning, the progeny of stars with mass $M\sim 5.5M_{\odot}$. 
This result opens up an important test of the strength hot bottom burning using
detailed near--IR (H and K bands) spectroscopic analysis of the oxygen--rich, 
high luminosity candidates found in the well defined region of the colour-colour plane.
This test is possible because the two stellar evolution codes we use predict very
different results for the surface chemistry, and the C/O ratio in particular, owing to 
their treatment of convection in the envelope and of convective boundaries during 
third dredge-up. The differences in surface chemistry are most apparent when the
model stars reach the phase with the largest infrared emmission. 

\end{abstract}


\section{Introduction}
Stars in the mass range $0.8M_{\odot} \lesssim M \lesssim 8M_{\odot}$ evolve through the Asymptotic 
Giant Branch (AGB)  \citep{becker80, iben82, iben83, lattanzio86}. This evolutionary phase is 
characterized by a series of thermal pulses, caused by the unstable ignition of helium in a
thin He--rich layer \citep{schw65, schw67}. Although the timescale of the AGB is 
relatively short in comparison with the previous phases of core nuclear burning, the richest 
nucleosynthesis and the strongest mass loss occurs during this phase of evolution. This means
that AGB stars are important for enriching the interstellar medium with the gas ejected 
from their surface layers and with the dust formed in their circumstellar envelopes.

Out of the stars that evolve through the AGB phase, those with initial masses above 
$\sim 4M_{\odot}$ exhibit an interesting behaviour and we will refer to these objects as
massive AGB stars. 
First, massive AGB stars experience a second dredge-up, which leads to a large increase
in the surface abundance of helium, which means that their ejecta is also strongly helium
rich.  Also, they experience Hot Bottom Burning (HBB), which occurs when the base of the 
convective mantle becomes sufficiently hot ($T_{\rm bce} > 30$MK) to ignite proton--capture 
nucleosynthesis. The results of this nuclear processing are rapidly transported to the 
surface by convective currents, thus rendering these objects possible efficient 
polluters of gas processed by CNO burning and more generally, by advanced proton--capture 
nucleosynthesis.

The strength of Hot Bottom Burning is highly dependent on the description of convection,
and especially so on the model used to determine the temperature gradient in regions unstable to 
convective motions. The extra luminosity from nuclear burning at the base of the convective envelope 
means that massive AGB stars deviate from the classic \citet{paczynski} relationship between 
core mass and luminosity \citep{renzini81, blocker91}\footnote{Garcia-Hernandez et al. (2009) 
observationally confirmed the extra HBB contribution to the luminosity in massive AGB stars 
in the Magellanic Clouds.}.
This was first shown by 
\citet{renzini81} in the 1980s, who also noted that the deviation depends on the efficiency 
of the convective model used.  \citet{dantona96}
showed that HBB conditions can be easily obtained when the Full Spectrum of Turbulence 
\citep[FST,][]{cm91} description of convection is used instead of the
traditional Mixing Length Theory (MLT). The study by \citet{vd05} importantly
showed that convection modelling is by far the most important ingredient affecting the
description of the evolution of massive AGB stars. Convection was shown to significantly
affect the duration of the  AGB phase, as well as the luminosity and the mass-loss rate
of massive AGB stars.

Owing to our poor knowledge of the efficiency of HBB, the predictive power of the
results of massive AGB stellar models is severely hampered.
Models experiencing strong HBB contaminate the interstellar medium with gas that
shows the signature of considerable CNO, Ne--Na and Mg--Al processing 
\citep{ventura08, ventura09, ventura13} and dust, mainly in the form of silicate 
particles \citep{paperI, paperII, paperIII, paperIV}. 
Conversely, in models with a less efficient HBB (and deeper third dredge-up, which mixes
primary carbon, oxygen and magnesium to the surface), the oxygen and magnesium content of 
the ejecta are not as strongly modified with respect to the initial abundances
\citep{karakas07, karakas10, karakas14}. These models can even become carbon
rich, where C/O $\ge 1$, at the end of the AGB phase, which means that solid
carbon grains are the main dust component formed in their winds \citep{fg06}.

All these uncertainties are a severe limitation to our understanding of the
role played by massive AGB stars in a number of astrophysical contexts. A few examples
of the importance of these objects include the possibility that they are the main actors in 
the formation of multiple populations in Globular Clusters \citep{ventura01}, that they
have an important contribution to the dust present at high redshift \citep{valiante09}, and
massive AGB stars are important for determining the chemical trends traced by stars in different
parts of the Milky Way, as shown by models of galactic chemical evolution 
\citep{romano10,kobayashi11}. 

These arguments stimulated us to start a comparative analysis, in an effort to
significantly improve our understanding of the main physical properties of
massive AGBs stars and particurly of the strength of HBB. The goal of this project is more
than  a mere comparison between results obtained with different stellar evolution models: 
we intend to test the robustness of the results obtained so far and to identify 
and suggest observations which could be relevant for discriminating among the different 
theoretical descriptions. 

In this work we propose to use results from Spitzer observations of the AGB population
of the Large Magellanic Cloud to allow a better understanding of their evolution properties. 
The LMC is an ideal laboratory for the study of AGB stars because they are 
relatively close \citep[$d \sim 50$kpc,][]{feast99}, with a low average reddening 
\citep[$E(B-V) \sim 0.075$,][]{schlegel98}. A growing body of observational 
data, based on dedicated photometric surveys, has been recently made available to the community: 
the Magellanic Clouds Photometric Survey \citep[MCPS,][]{zaritsky04}, the Two Micron All 
Sky Survey \citep[2MASS,][]{skrutskie06}, the Deep Near Infrared Survey of the Southern 
Sky \citep[DENIS,][]{epchtein94}, Surveying the Agents of a Galaxy's Evolution Survey 
\citep[SAGE--LMC with the {\it Spitzer} telescope,][]{meixner06}, and {\it HERschel} 
Inventory of the Agents of Galaxy Evolution \citep[HERITAGE,][]{meixner10, meixner13}. 
Additional data allowed to reconstruct the Star Formation History (SFH) of the LMC
\citep{harris09, weisz13}, and the age--metallicity relation \citep[AMR,][]{carrera08, 
piatti13}. These studies outline that a burst in star formation occurred 
$\sim 80-100$Myr ago \citep{harris09}, indicating a significant population
of massive AGB stars should be present in the LMC; these are the progeny of stars 
with initial mass $\sim5-6M_{\odot}$.

On the theoretical side, the recent investigations by \citet{flavia14a, flavia15} 
attempt an interpretation of the Spitzer observations of the LMC using evolution models of 
massive AGB stars,  
and a description of the dust formation mechanism in 
their winds. Dell'Agli et al. (2015) show that massive AGBs evolve into specific 
regions of the colour--colour and colour--magnitude planes obtained with the Spitzer bands,
separated by the zones populated by dust--free AGB stars and dusty carbon stars.
The agreement with the observations is remarkably good; 
however, the results obtained are extremely model dependent as the whole study used
AGB models based on the FST description of convection.

Here we make a step forward. We compare the results by Dell'Agli et al. (2015) with 
those based on different models of massive AGB stars.
Our scope is twofold: a) by comparing between models obtained with different prescriptions 
and codes, we aim to understand how stellar modelling uncertainties affect the dust
formation process in the winds of massive AGB stars, and how that affects their position
in the observational planes; 
and fix the uncertainties associated with the dust formation process in the winds of massive AGBs, 
and with the position occupied by these stars in the observational planes; 
b) we look for possible observations to be used to assess the strength of the HBB 
experienced by these sources.

The paper is organised as follows. The numerical and physical input to the AGB
evolution models and the dust formation process are given in Section \ref{inputs}. 
In Section~\ref{agb} we describe the main physical features of the massive AGB models, 
whereas the properties of the dust formed in their winds is addressed in Section \ref{dust}. 
In Section \ref{spitzer} we discuss the evolution of these stars in the observational
planes obtained with the Spitzer bands.

\section{Numerical and physical inputs}
\label{inputs}
The massive AGB star population in the LMC formed during the burst in the SFH that occurred
$\sim 80-100$Myr ago \citep{harris09}. 
Half of the stars formed in that epoch have a metallicity $Z=8\times 10^{-3}$, with
an additional lower metallicity component, with $Z=10^{-3}$.
Taking into account the evolution times of stars of
intermediate mass, we deduce that most of the massive AGB stars observed today in the LMC
are the descendants of stars with initial mass $M \sim 5.5M_{\odot}$. Based on these
arguments, we will focus in the following on stellar models of mass $M = 5.5M_{\odot}$
and metallicity $Z=8\times 10^{-3}$. We are interested to the behaviour of dusty, obscured,
oxygen--rich
sources and for this reason we neglect the lower-$Z$ component. These low-metallicity objects
have a smaller silicon abundance, which leads to lower dust production rates \citep{paperIV}
during the oxygen--rich phase.

\subsection{Stellar evolution models}
The models presented in this work were calculated with the ATON code \citep{ventura98} 
and with the Monash version of the Mount Stromlo Stellar Structure Program \citep{frost96}. 
We will refer to these models as the ATON and MONASH models, respectively.

The interested reader is referred to the papers by \citet{ventura13} and 
\citet{karakas10} for a detailed discussion
of the numerical and physical inputs used to calculate the evolutionary sequences and for
an exhaustive description of the chemical and physical properties of the AGB evolution of
these stars. Both models were evolved until the almost complete 
loss of the convective envelope.

The main differences between the two sets of models is in the treatment 
of convection and the description of mass loss. In the ATON code the convective instability 
is described by means of the FST model, whereas in the MONASH case the traditional MLT 
treatment is used. The mass-loss rate in the ATON case is determined via the \citet{blocker95}
treatment, whereas the MONASH models adopt the \citet{VW93} mass-loss prescription.

\subsection{Dust formation in the winds of AGB stars}
\label{dustmodel}
The growth of dust particles is calculated with a simple model for the stellar wind. 
This is based on the pioneering explorations by the Heidelberg group \citep{gs85, gs99, 
fg01, fg02, fg06, zhukovska08} and was extensively used in previous works by our group
\citep{paperI, paperII, paperIII, paperIV}, as also in works of other researchers
\citep{nanni13a, nanni13b, nanni14}.
 
The outflow is assumed to expand symmetrically from the stellar surface, with an initial
velocity of $1$ km s$^{-1}$. The description of the wind is given by solving two differential
equations, which describe the radial variation of the gas velocity and of the optical depth,
$\tau$:

\begin{equation}
v{dv\over{dr}}=-{GM_*\over r^2}(1-\Gamma),
\label{eqvel}
\end{equation}

\begin{equation}
{d\tau\over{dr}}=-k\rho\left({R_*\over r}\right)^{2} ,
\end{equation}

where $M_*$ and $R_*$ are the stellar mass and luminosity, $\rho$ is the density of the
gas, $k$ is the extinction coefficient and $\Gamma$ is given by the expression

\begin{equation}
\Gamma={kL_*\over{4\pi cGM_*}},
\label{eqgamma}
\end{equation}

with $L_*$ indicating the luminosity of the star.

The above equations are completed by the mass conservation equation, for density, and the
relationship governing the radial variation of temperature as a function of the effective 
temperature of the star:

\begin{equation}
\rho={\dot M\over{4\pi r^2 v}},
\label{eqrho}
\end{equation}

\begin{equation}
T^4={1\over 2}T_{eff}^4\left[1-\sqrt{1-\left({R_*\over r}\right)^2}\right]+{3\over 2}\tau.
\label{eqt}
\end{equation}

$\Gamma$ is the key quantity for the dynamics of the wind: the expanding gas can be
accelerated only when the condition $\Gamma > 1$ is reached. Eq. \ref{eqgamma} shows
that the acceleration of the wind is favoured by large luminosities, as a consequence of
the large radiation pressure determined by large values of $L_*$. A high $k$ can also
produce a large acceleration of the wind. The value of $k$ depends on the number
density and the size of the dust particles of the various species formed. The growth of
the dimension of the various dust particles forming in the wind is described by means
of additional equations, giving the growth rate of each species as a function of the
local values of density and temperature and of the surface abundance of the elements
relevant for the formation of each dust species. 

In oxygen--rich environments (C/O below unity ) we consider the formation of alumina dust 
($Al_2O_3$), silicates and solid iron. The relevant elements for the formation of these 
dust species are aluminium, silicon and iron. When $C/O > 1$ we follow the formation of 
solid carbon grains, silicon carbide and solid iron; in this case the key--elements are
carbon, silicon and iron.

The whole set of equations describing the dust formation process is described and commented
in details in \citet{paperI, paperII, paperIII, paperIV}.

\subsection{Synthetic spectra}
\label{spectramodel}
For each point of the evolutionary sequence we compute the magnitudes in the various
Spitzer bands by calculating the synthetic spectra. This is done in two steps:

\begin{itemize}
\item{The results from stellar evolution modelling, particularly the values of mass,
luminosity, effective temperature, mass-loss rate and surface chemical composition, are
used to find out the dust species formed in the wind, the size of the dust grains and the
optical depth (here we use the value at $10\mu$m, $\tau_{10}$).}

\item{By means of the code DUSTY \citep{dusty} we calculate the synthetic spectra of each 
selected point along the evolutionary sequence; the magnitudes in the various bands are 
obtained by convolution with the appropriate transmission curves.}
\end{itemize}

The code DUSTY needs as input parameters the effective temperature of the star, the
radial profile of the density of the gas and the dust composition of the wind, in terms
of the percentage of the various species present and of the the size of the dust particles
formed. All these quantities are known based on the results of stellar evolution and of
the description of the wind. 

In the recent investigation by Dell'Agli et al. (2015) we give the details on the 
way DUSTY is used to determine the synthetic spectra both of oxygen--rich stars and of
carbon stars.

\section{The role of Hot Bottom Burning in the evolution of massive AGBs}
\label{agb}
The main features of the evolution of stars of low and intermediate mass along the
AGB phase have been thoroughly discussed in the literature, together with the 
uncertainties associated to the various physical inputs adopted. We address the
interested reader to the exhaustive reviews on this argument by \citet{herwig05}
and \citet{karakas14}. 

In massive AGB models, two mechanisms may potentially alter
the surface chemical composition: Hot Bottom Burning and Third Dredge Up (TDU).
In the first case the surface chemistry changes according to the equilibrium
abundances of proton--capture nucleosynthesis, which in turn depends on the
temperature at the base of the convective envelope.
The ignition of HBB determines the decrease in 
the surface carbon in favour of nitrogen; when the HBB is strong (with temperatures
above $\sim 70$ MK) oxygen and magnesium are destroyed and sodium and aluminium 
are produced. TDU is the inward penetration of the surface
convection after each thermal pulse down to layers contaminated by $3\alpha$
nucleosynthesis. The main modification of the surface chemical composition by
TDU is the gradual increase in the carbon content, although both oxygen
and magnesium can also be dredged up as a consequence of partial helium burning \citep{herwig00,karakas03}.

The ATON and MONASH codes have been written and developed independently from each other;
they differ in the numerical structure and in the description of some physical mechanisms
relevant for the evolution on the AGB, primarily convection and mass loss. We
therefore expect that the differences among the results obtained provide a valuable 
indication of the uncertainties affecting the evolution of massive AGB stars.

To this aim, we compare the results concerning the evolution of two models of initial
mass $M=5.5M_{\odot}$ and metallicity $Z=8\times 10^{-3}$. The reason behind this choice
is partly that this mass is within the range of masses experiencing HBB; additionally,
on the basis of the arguments presented in section \ref{inputs}, we know that in the LMC
this is the typical mass evolving today as a massive AGB star.
 
\begin{figure*}
\begin{minipage}{0.48\textwidth}
\resizebox{1.\hsize}{!}{\includegraphics{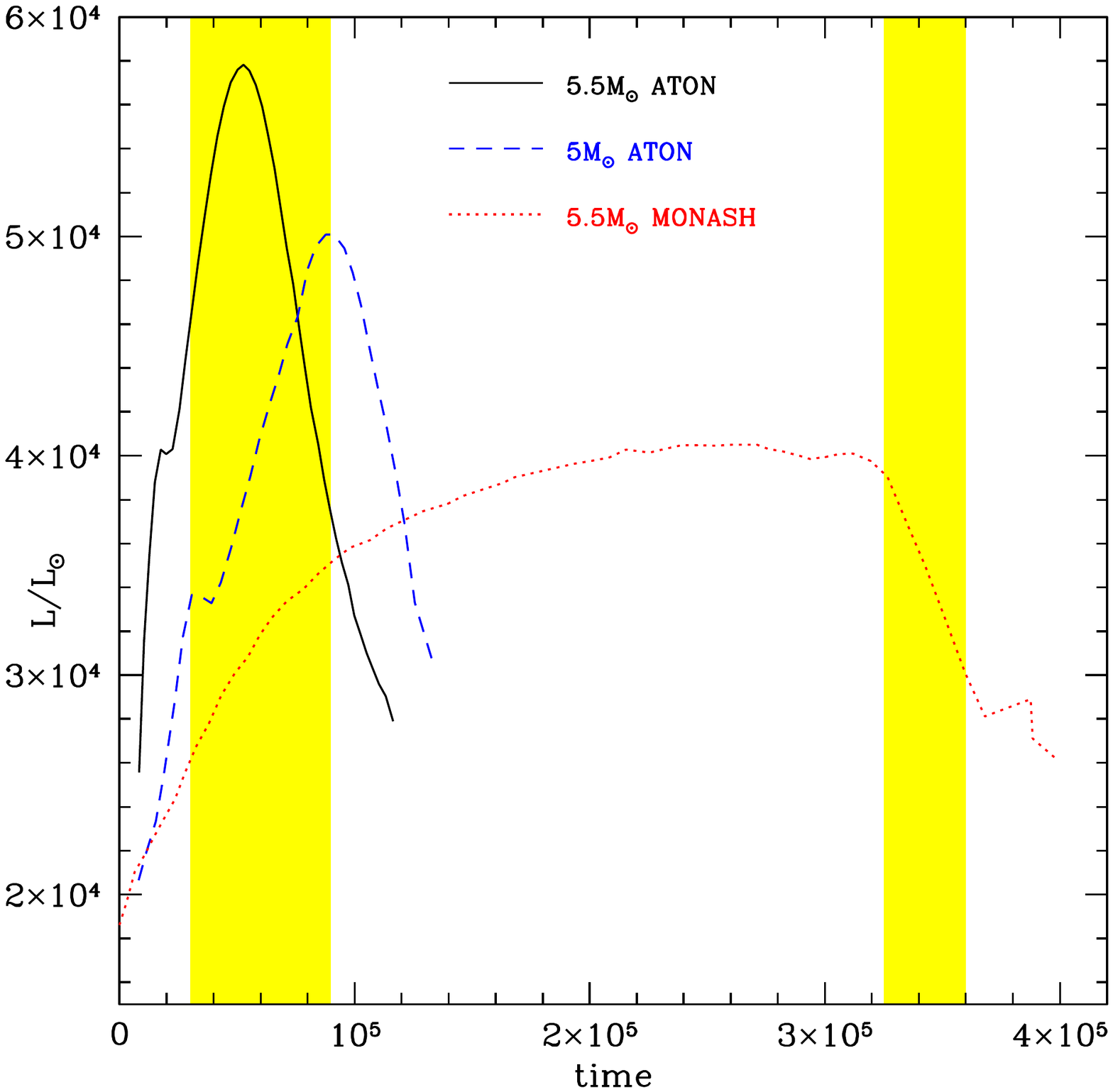}}
\end{minipage}
\begin{minipage}{0.48\textwidth}
\resizebox{1.\hsize}{!}{\includegraphics{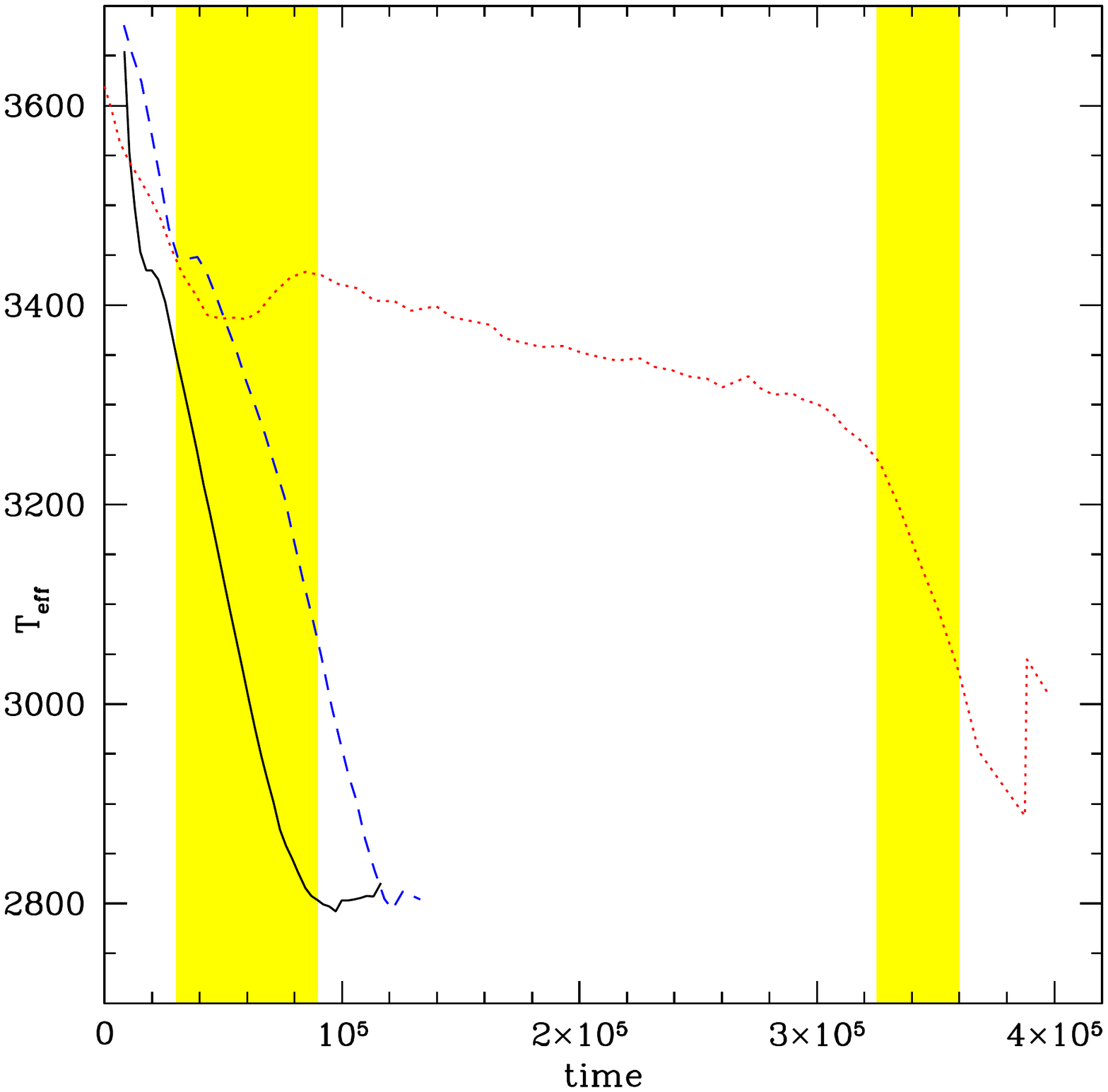}}
\end{minipage}
\vskip-70pt
\begin{minipage}{0.48\textwidth}
\resizebox{1.\hsize}{!}{\includegraphics{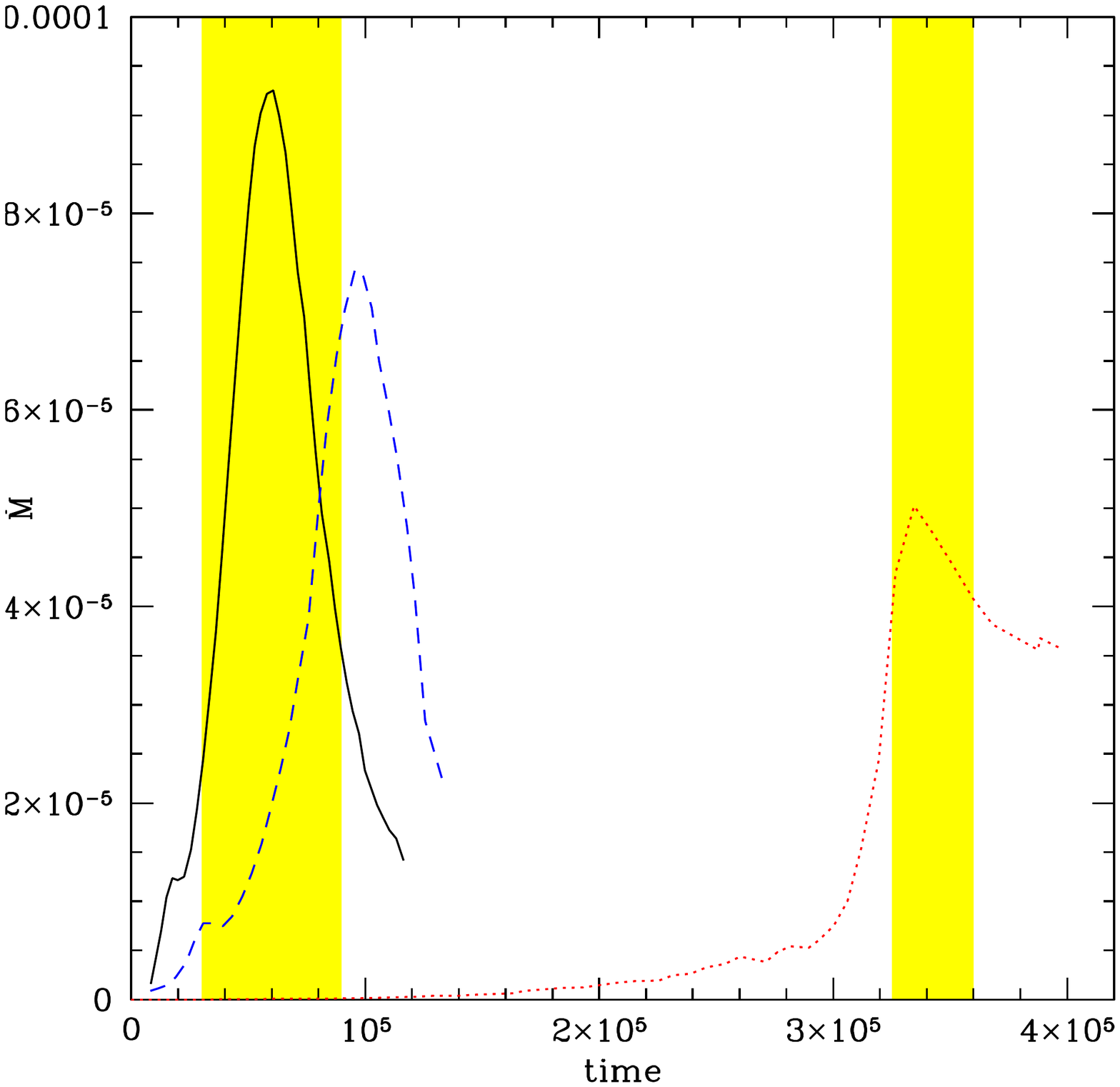}}
\end{minipage}
\begin{minipage}{0.48\textwidth}
\resizebox{1.\hsize}{!}{\includegraphics{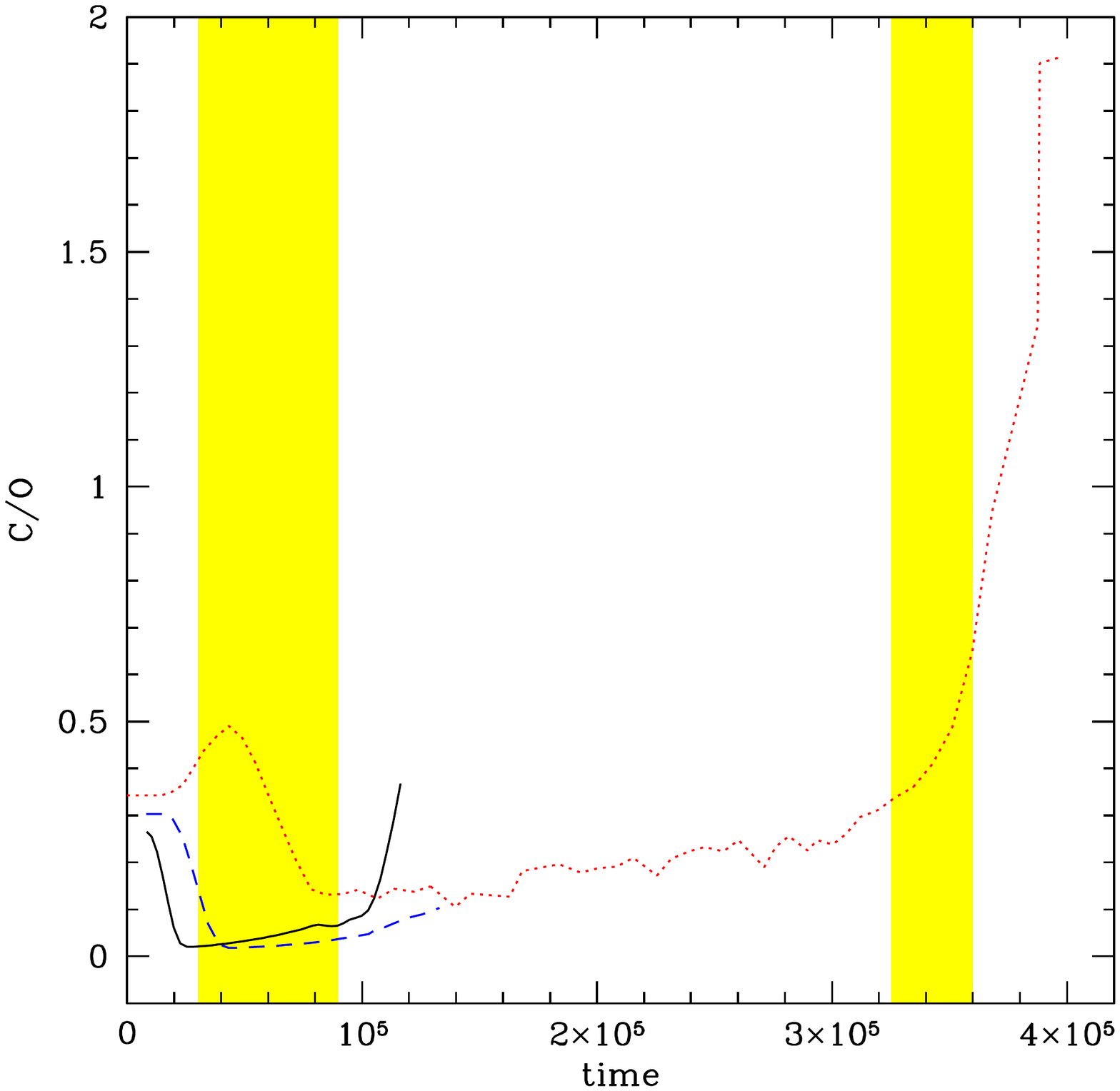}}
\end{minipage}
\vskip-30pt
\caption{The evolution of luminosity (left--top panel), effective temperature
(right--top), mass-loss rate (left--bottom) and surface C/O ratio (right--bottom)
of massive AGB models calculated with the the ATON and MONASH stellar evolution codes.
Times are counted from the beginning of the AGB phase. The shaded regions indicate
the phase of highest mass loss rate and strongest infrared emission for the ATON
and MONASH models of initial mass $5.5M_{\odot}$.}
\label{fhbb}
\end{figure*}


Fig. \ref{fhbb} shows the temporal variation of the main
physical and chemical quantities of the ATON and MONASH models during the AGB phase.
We show the evolution of the luminosity, the mass-loss rate, the effective temperature, 
and the surface C/O ratio.

Common features in the behaviour of the two sequences are the gradual increase in the
luminosity $L$, characterizing the first part of the AGB evolution, due to an increase in the
core mass, and the decline of $L$ in the final AGB phases, a consequence of the progressive 
shutting down of HBB. 
The rate at which mass loss occurs follows a similar trend, reaching the highest
value in the middle of the AGB phase.
The surface regions become cooler and cooler as the total mass of the star diminishes: 
the effective temperature decreases from the initial value of $T_{\rm eff} \sim 3600$K to 
$T_{\rm eff} \sim 2800$K in the latest stages. 

Other than these similarities, we see from Fig. \ref{fhbb} that the two models exhibit 
considerable differences.

First, the ATON models evolve to larger luminosities. The $5.5M_{\odot}$ ATON model reaches
a maximum luminosity of $\sim 6\times 10^4L_{\odot}$, whereas the corresponding MONASH
model reaches $L \sim 4\times 10^4L_{\odot}$. To show that the difference between
the two models is intrinsic, we also show the evolution of the ATON $5M_{\odot}$ model, 
which, though evolving on a core of smaller mass, reaches a higher luminosity 
($L \sim 5\times 10^4L_{\odot}$) than the MONASH case. This difference is not surprising,
considering the different description of convection adopted in the two cases. 
\citet{vd05} clearly showed that the use of the FST model leads to stronger HBB conditions and
to higher luminosities. We note in Fig. \ref{fhbb} that in the ATON case, owing to an early
ignition of HBB, the luminosity starts to increase very fast since the early AGB phases, 
at odds with the MONASH model, in which the increase in the overall flux is more gradual.

The rate at which mass is lost reflects the differences among the luminosities. 
ATON models experience larger mass-loss rates, which reach $\dot M \sim 10^{-4} M_{\odot}/yr$ in
the phase of maximum luminosity. The MONASH model experience a weaker mass loss:
$\dot M < 5\times 10^{-5} M_{\odot}/yr$ during the whole AGB phase. The mass-loss rate
experienced by the ATON models reflects the strong variation in the luminosity. This is because 
the \citet{blocker95} description of mass loss is much more sensitive to luminosity 
than the \citet{VW93} recipe.

The strong mass-loss rates experienced by the ATON models has an important consequence
for the duration of the AGB phase. The duration of the AGB phase in the MONASH model is 
$\sim 4\times 10^5$ yr, a factor of 2.6 longer than the ATON case, which has a duration
of less than $\sim 1.5\times 10^5$ yr.
%

The evolution of the surface chemistry also shows considerable differences between the two
models. Here we focus on the C/O ratio, a valuable indicator of the strength of HBB and of 
the extent of the Third Dredge--Up.
In the ATON models (see Fig. \ref{fhbb}) HBB favours a strong depletion of the surface C/O
after a few $10^4$yr, owing to the destruction of the surface carbon, via proton fusion.
The C/O ratio remains below $\sim 0.05$ for the rest of the AGB evolution, because
the carbon transported to the surface by TDU is immediately destroyed by HBB. Only in the very 
final evolutionary phases, when HBB is practically quenched by the general cooling 
determined by the gradual loss of the convective envelope, the surface C/O increases. 
The MONASH model experiences a weaker HBB.
In the competition between TDU and HBB, the latter prevails only in the period
$2-4\times 10^5$yr, when the C/O ratio drops to 0.15. Following HBB shutting down, the C/O 
ratio gradually  increases and in the very final phases the model becomes carbon rich.

\section{Dust from massive AGB stars}
\label{dust}
A common feature of the ATON and MONASH models is that
for most of the AGB evolution massive AGB stars evolve as oxygen--rich stars. Only during
the final few thermal pulses does the MONASH model reach the carbon star stage.

The amount of dust formed in the winds of oxygen--rich AGBs is affected by the
strength of the HBB experienced \citep{paperII}: models suffering stronger HBB evolve
at larger luminosities, loose their envelope with a higher rate, which, in turn, favours
the growth of dust particles in the wind. Based on the differences between the physical
properties of the ATON and MONASH models outlined in the previous section, we focus now
on the description of how these reflect into the dust formation process expected in the two 
cases.

In oxygen--rich environments the most stable
species is alumina dust, which forms at temperatures $T \sim 1500$K, at a distance from
the surface of the star of the order of $\sim 2-3$ stellar radii \citep{flavia14b}. 
Additional species forming are silicates, in the form of olivine, pyroxene
and quartz \citep{fg06}. The silicates are less stable than alumina dust and only form in more
external regions ($d\sim 7-10R_*$), at temperatures $T \sim 1100$K. Solid iron is even
less stable, thus it forms in even more external and cooler layers, in smaller quantities.

\begin{figure*}
\begin{minipage}{0.47\textwidth}
\resizebox{1.\hsize}{!}{\includegraphics{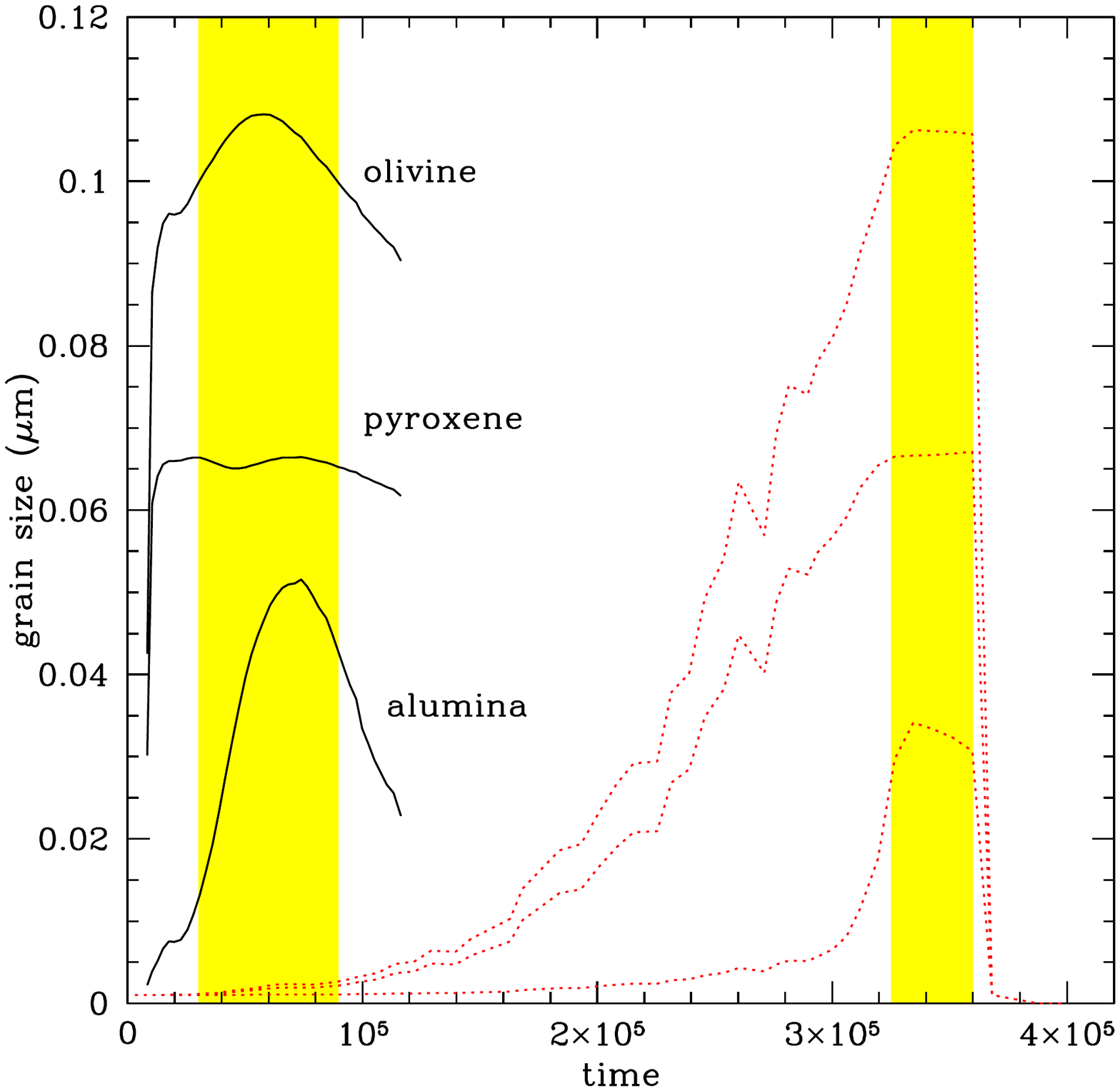}}
\end{minipage}
\begin{minipage}{0.47\textwidth}
\resizebox{1.\hsize}{!}{\includegraphics{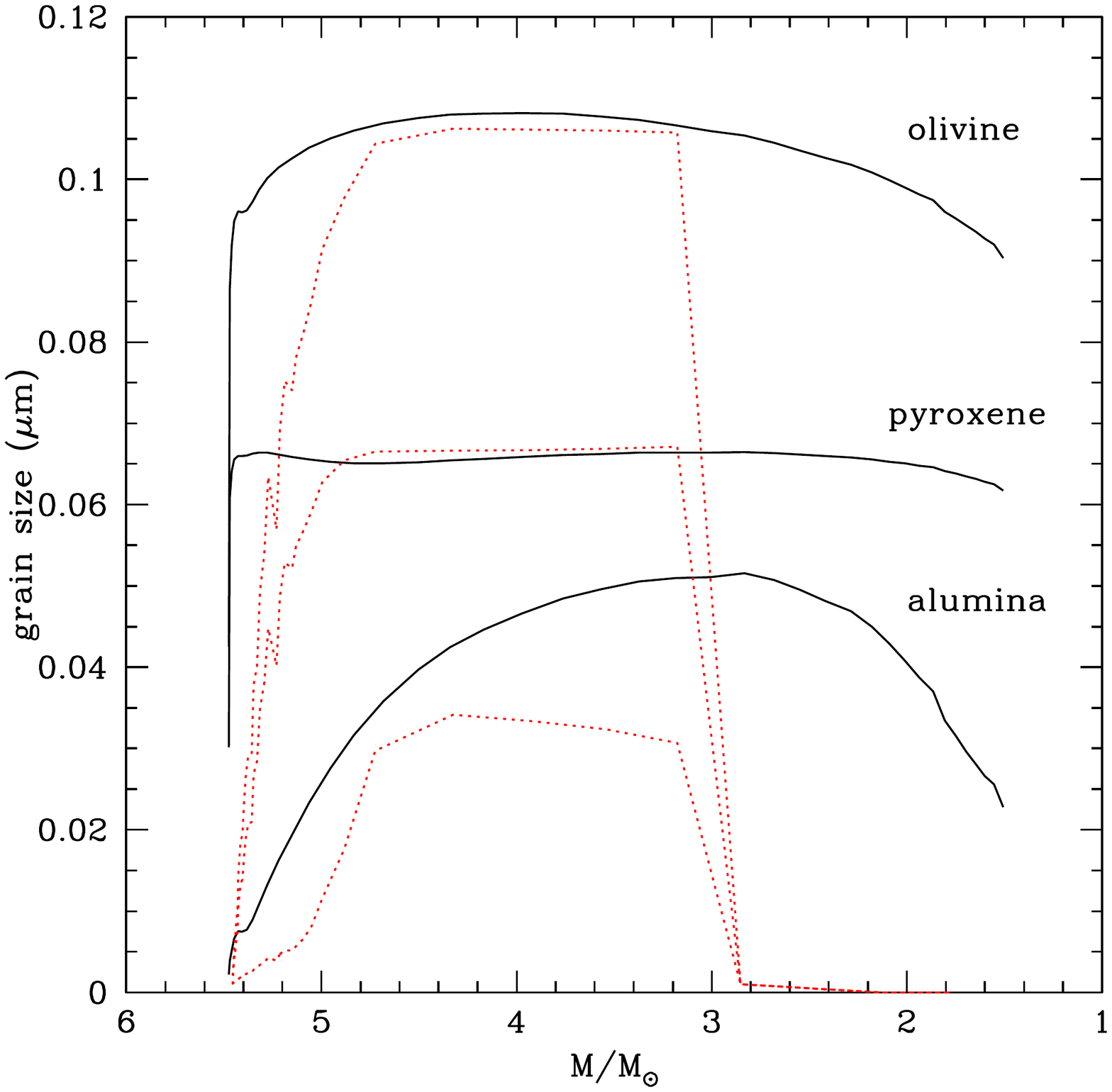}}
\end{minipage}
\vskip-40pt
\caption{Evolution of the size of the olivine, pyroxene and alumina dust grains
formed in the wind of the $5.5M_{\odot}$ models shown in Fig. \ref{fhbb} 
during the AGB phase. The variation of the grain size as a function of the time
counted from the beginning of the AGB phase (left panel) and of the current mass of the
star (right panel) are shown. The two sets of tracks indicate results from ATON (solid) and
MONASH (dotted) codes. The shaded regions in the left panel have the same meaning as
in Fig. \ref{fhbb}.
}
\label{fsize}
\end{figure*}

Alumina dust is transparent to electromagnetic radiation. This means that the
wind is barely (or not at all) accelerated by the formation of $Al_2O_3$ grains.
Therefore the wind enters the region of silicates formation with a velocity close to the
initial velocity. The main consequence is that the formation of silicates is 
not affected by the amount of alumina dust formed.

The presence of silicate grains has a much stronger influence on the dynamics of the wind:
the large values of the extinction coefficient, $k$, provokes a strong acceleration
of the gas particles, owing to the increase of $\Gamma$ (see Eqs. \ref{eqvel} and 
\ref{eqgamma} in Section \ref{inputs} and the discussion on the role of $\Gamma$ on the
acceleration of the wind).

Fig. \ref{fsize} shows the evolution of the size of the various dust particles formed 
during the AGB phase for the ATON and MONASH models, discussed in the previous section. 
The ATON $5M_{\odot}$ is not shown here, as we have seen that it shares many similar properties 
to the more massive models.  For clarity we show only the size of the three 
most abundant species formed, i.e., olivine, pyroxene and alumina dust. We see that the 
dust properties of the two
models during the highest luminosity phase are very similar. In both cases we find  
that the central star is surrounded by an optically thick dust layer, dominated by
olivine grains, with size slightly above $0.1\mu$m and pyroxene particles of smaller
size, with typical dimension $\sim 0.06-0.07 \mu$m. The only differences found is in the
properties of the alumina dust formed. In the ATON model the $Al_2O_3$ grains reach
a larger size ($\sim 0.05 \mu$m) compared to the MONASH case ($\sim 0.03-0.04 \mu$m).
The reason for the larger grain growth in the ATON models is explained below.

\begin{figure*}
\begin{minipage}{0.48\textwidth}
\resizebox{1.\hsize}{!}{\includegraphics{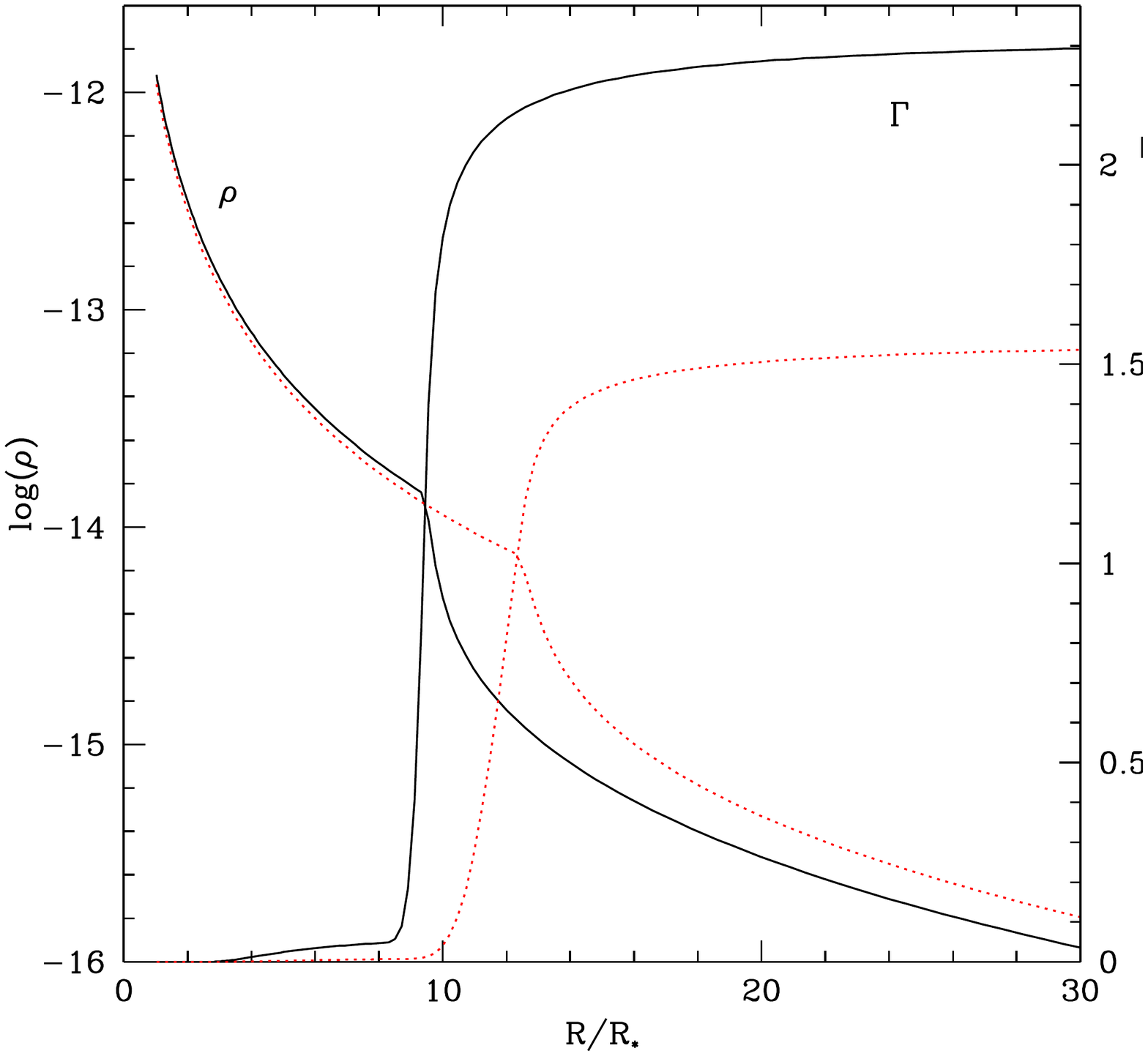}}
\end{minipage}
\begin{minipage}{0.48\textwidth}
\resizebox{1.\hsize}{!}{\includegraphics{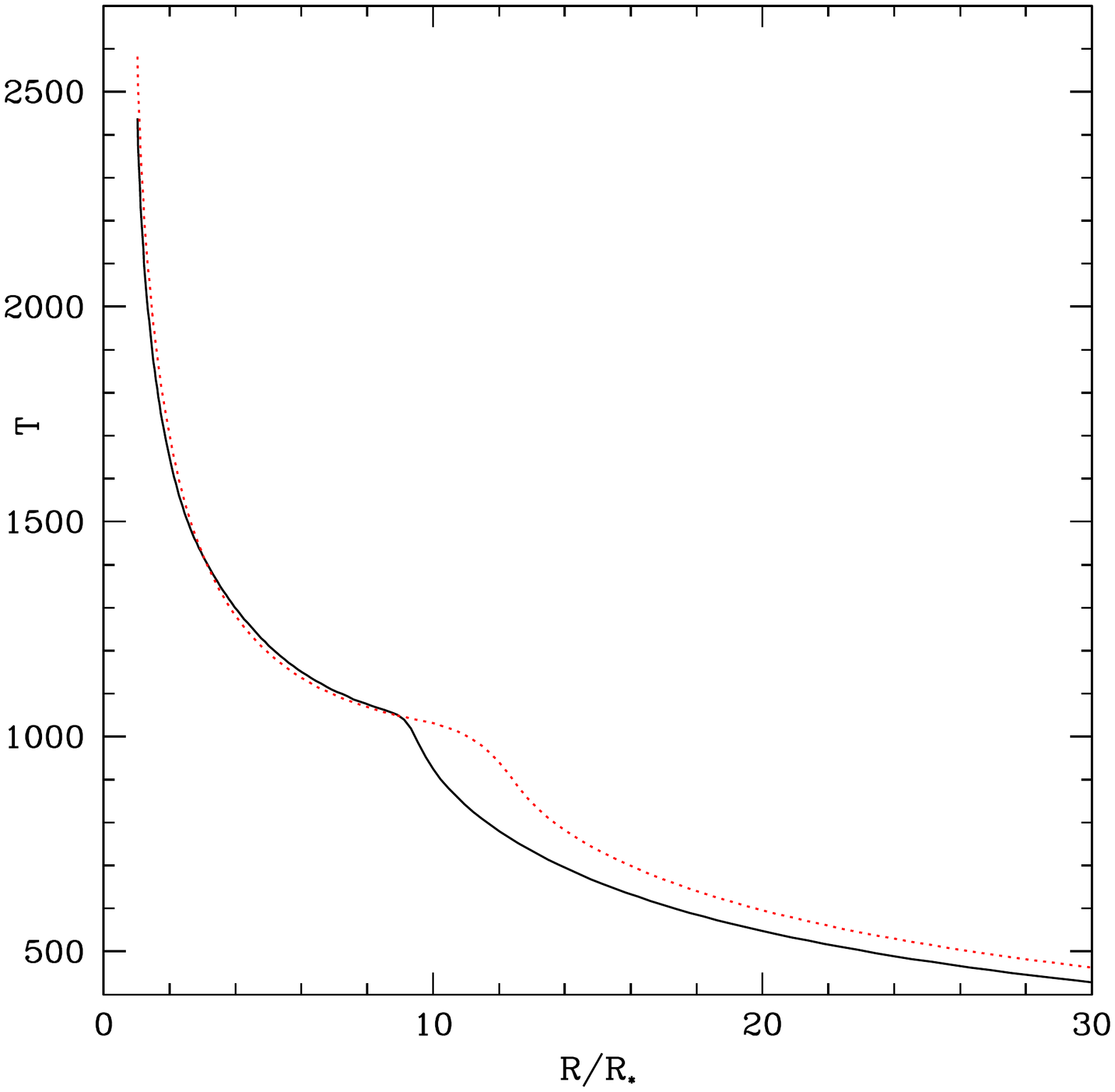}}
\end{minipage}
\vskip-70pt
\begin{minipage}{0.48\textwidth}
\resizebox{1.\hsize}{!}{\includegraphics{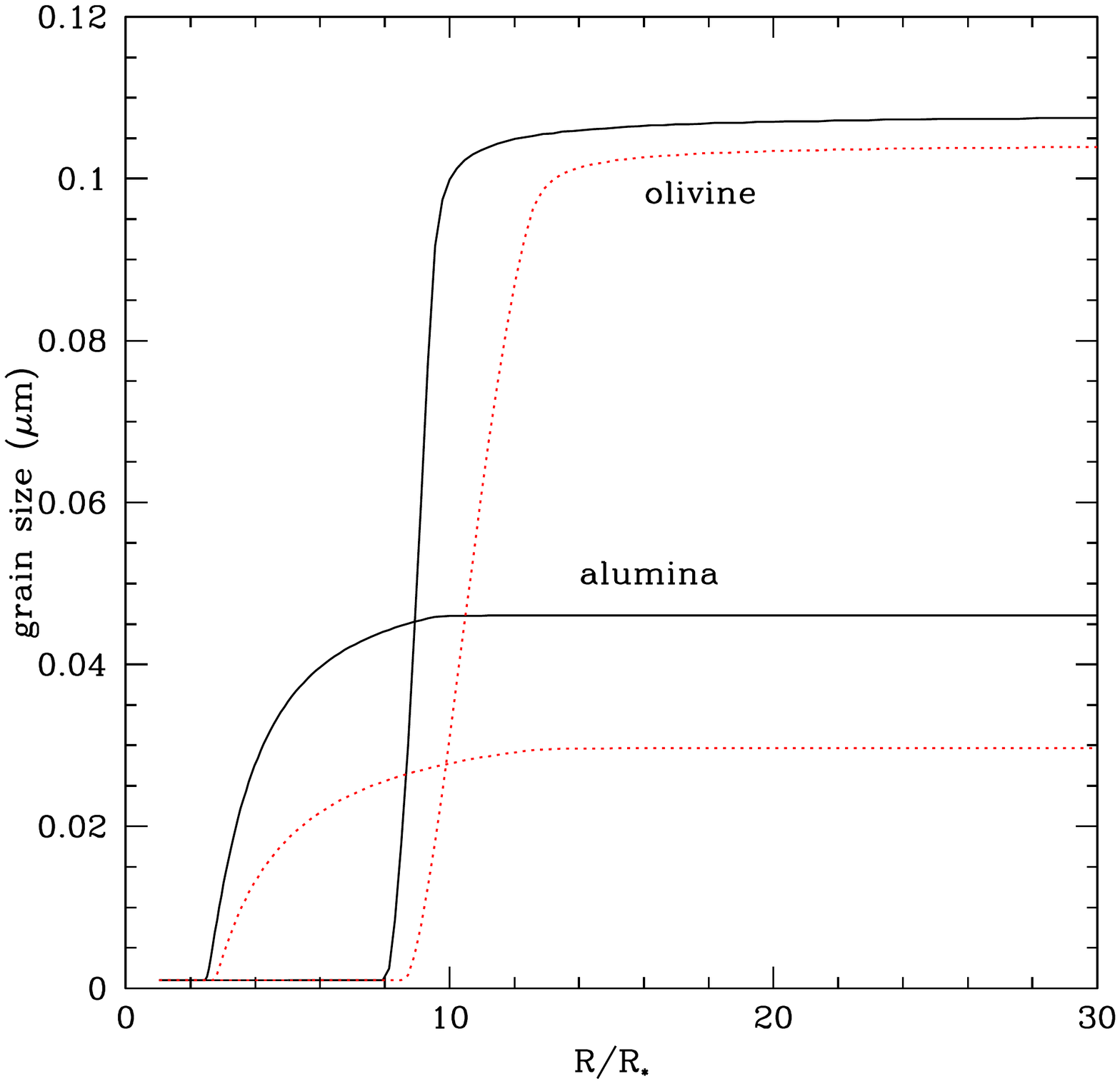}}
\end{minipage}
\begin{minipage}{0.48\textwidth}
\resizebox{1.\hsize}{!}{\includegraphics{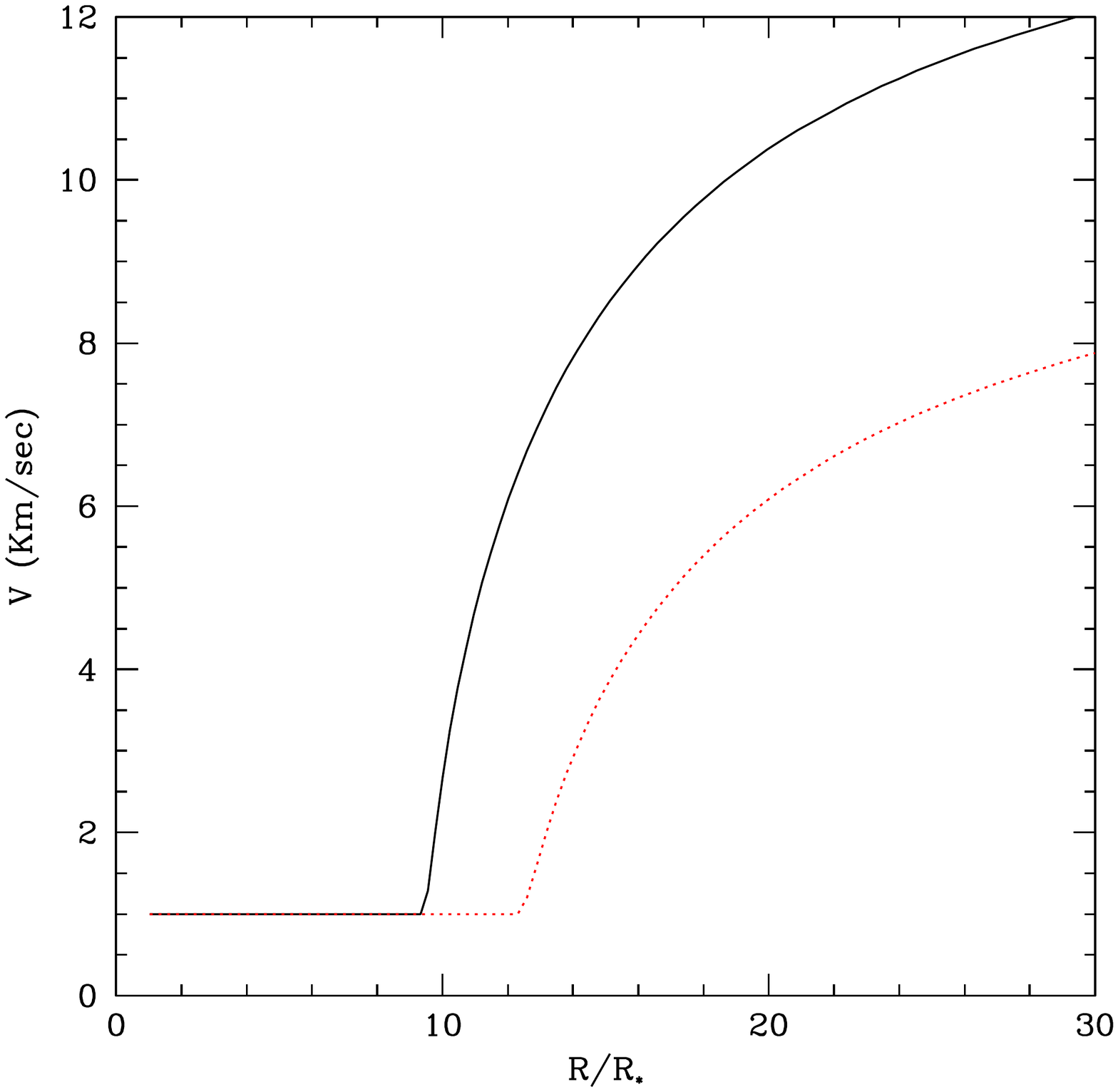}}
\end{minipage}
\vskip-30pt
\caption{The structure of the wind of the ATON (solid) and MONASH (dotted)
models during the phase of highest mass loss rate. The quantities shown are the radial
profiles of density and $\Gamma$ (left--top panel), temperature (right--top), size of 
the grains formed (left--bottom) and velocity of gas (right--bottom). Distances are
measured from the centre of the star and are expressed in units of stellar radii.
}
\label{fwind}
\end{figure*}

This result is at first surprising, given the differences in the physical AGB evolution 
of the two models. To understand the reason for the similarities found, we need to analyze 
the dynamics of the wind and the coupling with the dust formation process.

To this aim, we show in Fig. \ref{fwind} the radial variation of the thermodynamic quantities,
including the size of alumina dust and olivine grains formed, the velocity of the wind and of
$\Gamma$. This figure refers to the ATON and MONASH models during the respective phases 
of highest mass loss rate.

In the previous section we have seen that the ATON model experiences a higher mass-loss
rate. This favours a higher density in the wind (see Eq. \ref{eqrho}) and a more efficient
formation of dust, owing to a larger availability of gas molecules in the wind. On the 
other hand the FST models are more luminous and evolve on more expanded configurations.
This has no effect on the temperature stratification, which depends on $r/R_*$ 
(see Eq. \ref{eqt}), but it affects the density, which, based on Eq. \ref{eqrho}, scales 
as $1/r^2$. This effect partly compensates for the higher $\dot M$ of the ATON model,
decreasing the density gap with the MONASH case. The production of alumina dust is higher
in the ATON case, the size of the $Al_2O_3$ grains growing faster as the wind expands
outwards. 

As discussed above, formation of alumina dust is not sufficient
to accelerate the wind. The profile of $\Gamma$ remains practically flat (we note only
a modest increase of $\Gamma$ in the ATON case, but still with $\Gamma<0.1$) in the whole 
region internal to the zone where the formation of silicates occurs.

When entering the region where formation of silicates begins, the ATON wind is denser
than in the MONASH case but the difference is below a
factor $\sim 2$, as explained above. Unlike with alumina dust, the formation of silicates provokes a
strong acceleration of the wind, owing to the effects of radiation pressure. This can
be clearly seen in the increase in the values of velocity and of $\Gamma$, accompanying the
increase in the size of olivine grains. The acceleration experienced by the MONASH wind
is smaller but this indirectly favours the formation of silicates, because the condensation
zone is wider. This is the reason why the size reached by olivine grains in the MONASH 
model is only slightly smaller than in the ATON case. 

The results shown in Fig. \ref{fsize} are of extreme importance for the studies focused
on dust production around massive AGB stars. Despite the fact that our comparison identified
significant differences in several physics aspects of AGB evolution between the ATON and
MONASH models, the results in terms of the dust formation process are much more
homogeneous.  During the phase when the models reach the highest rate of mass loss, they are both 
surrounded by two dusty layers: an internal region, populated by alumina dust grains 
(with typical dimension of $\sim 0.05 \mu$m), and a more external zone, with silicates 
grains of $\sim 0.1 \mu$m size. These results are rather robust at the metallicity 
of the LMC ($Z = 8\times 10^{-3}$) and independent of the details of AGB modelling.

This is a welcome result for the reliability of this kind of investigation. The main
criticism put forward for this description is that the mass-loss rate is assumed as
a boundary condition, rather than being deduced on the basis of the amount of dust formed.
However, these results confirm that for massive AGB stars the dust formation
process is not strongly dependent on the mass-loss rate assumed.

In terms of the overall dust mass produced by these stars, we find that models
experiencing strong HBB produce more silicates and alumina dust. In the ATON case 
the mass of silicates and of $Al_2O_3$ produced are, respectively, 
$M_{sil}=3\times 10^{-3}M_{\odot}$ and $M_{Al_2O_3}=2\times 10^{-4}M_{\odot}$,
whereas the MONASH model gives $M_{sil}=1.7\times 10^{-3}M_{\odot}$ and 
$M_{Al_2O_3}=4.2\times 10^{-5}M_{\odot}$. In the latter
case we also have production of carbon dust, with mass $M_{C}=4.5\times 10^{-4}M_{\odot}$.
The reason for this differences can be understood based on the right panel of
Fig. \ref{fsize}. In the ATON case silicate particles are produced during the whole
phase of mass loss, whereas in the MONASH model this is
restricted to the phases when the rate of mass loss attains its largest values: the 
production of silicates is modest in the initial AGB phases and is null in the final 
part of the evolution, when the C/O ratio exceeds unity.

A definitive confirm to these findings can be obtained only on the basis of
radiation-hydrodynamical models.
The analysis by \citet{hofner08} and \citet{bladh12} confirmed that iron--free silicates, 
particularly $Mg_2SiO_4$, are viable wind--drivers in O--rich stars, owing to the
significant contribution of scattering to their extinction coefficients. This
self--consistent method, applied to the winds of stars of smaller mass 
($M\sim 1M_{\odot}$) and lower luminosity ($L < 10^4L_{\odot}$) than those of
interest here, showed that silicate particles of size in the range $\sim 0.1\ - 1 \mu$m
can potentially accelerate the wind. A similar approach is needed for the 
massive AGB stars examined in this work to assess whether the typical dimension
of $\sim 0.12\mu$m, found in our case, is sufficient to favour radiative acceleration of
the wind.

\begin{figure*}
\begin{minipage}{0.47\textwidth}
\resizebox{1.\hsize}{!}{\includegraphics{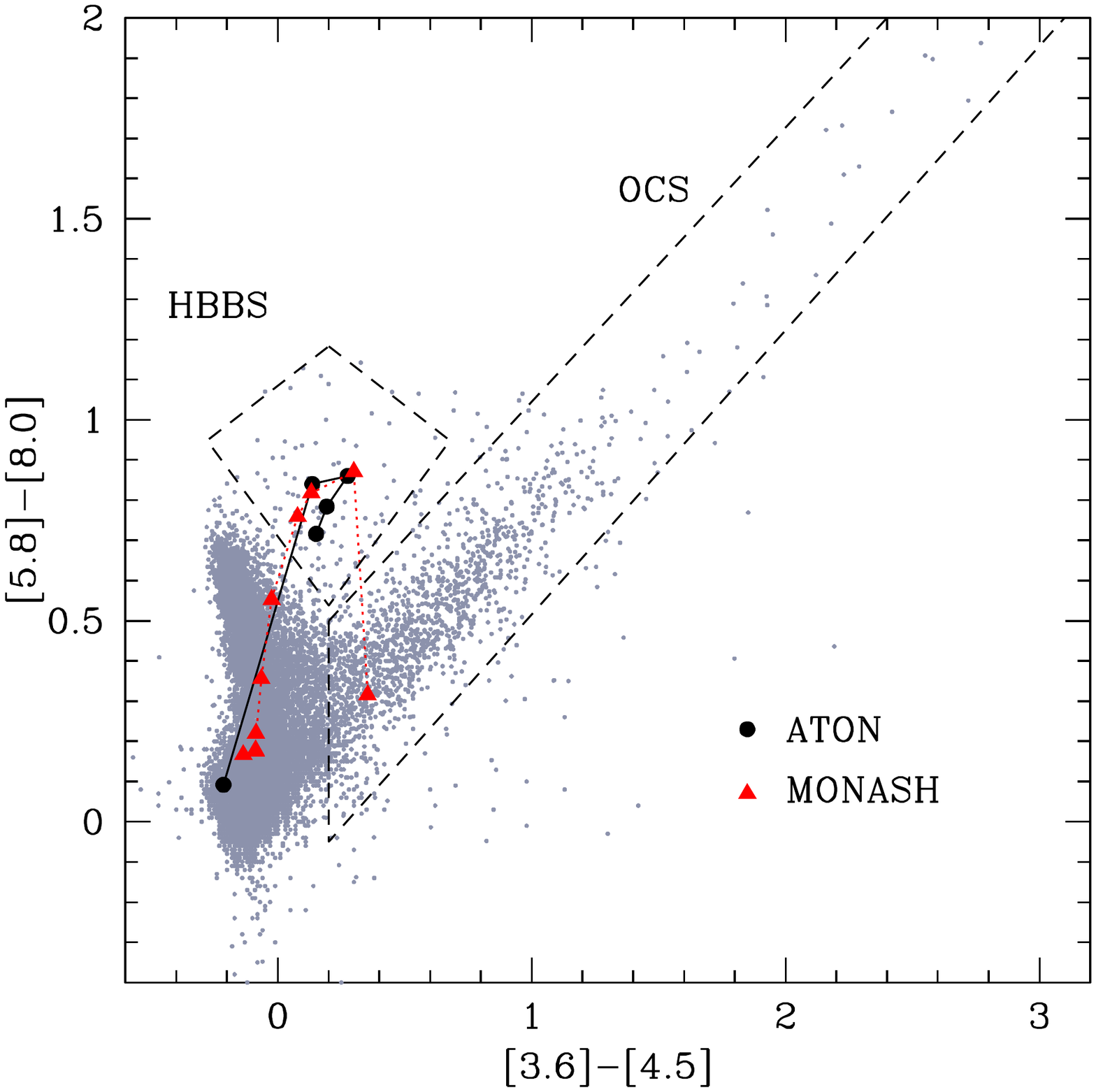}}
\end{minipage}
\begin{minipage}{0.47\textwidth}
\resizebox{1.\hsize}{!}{\includegraphics{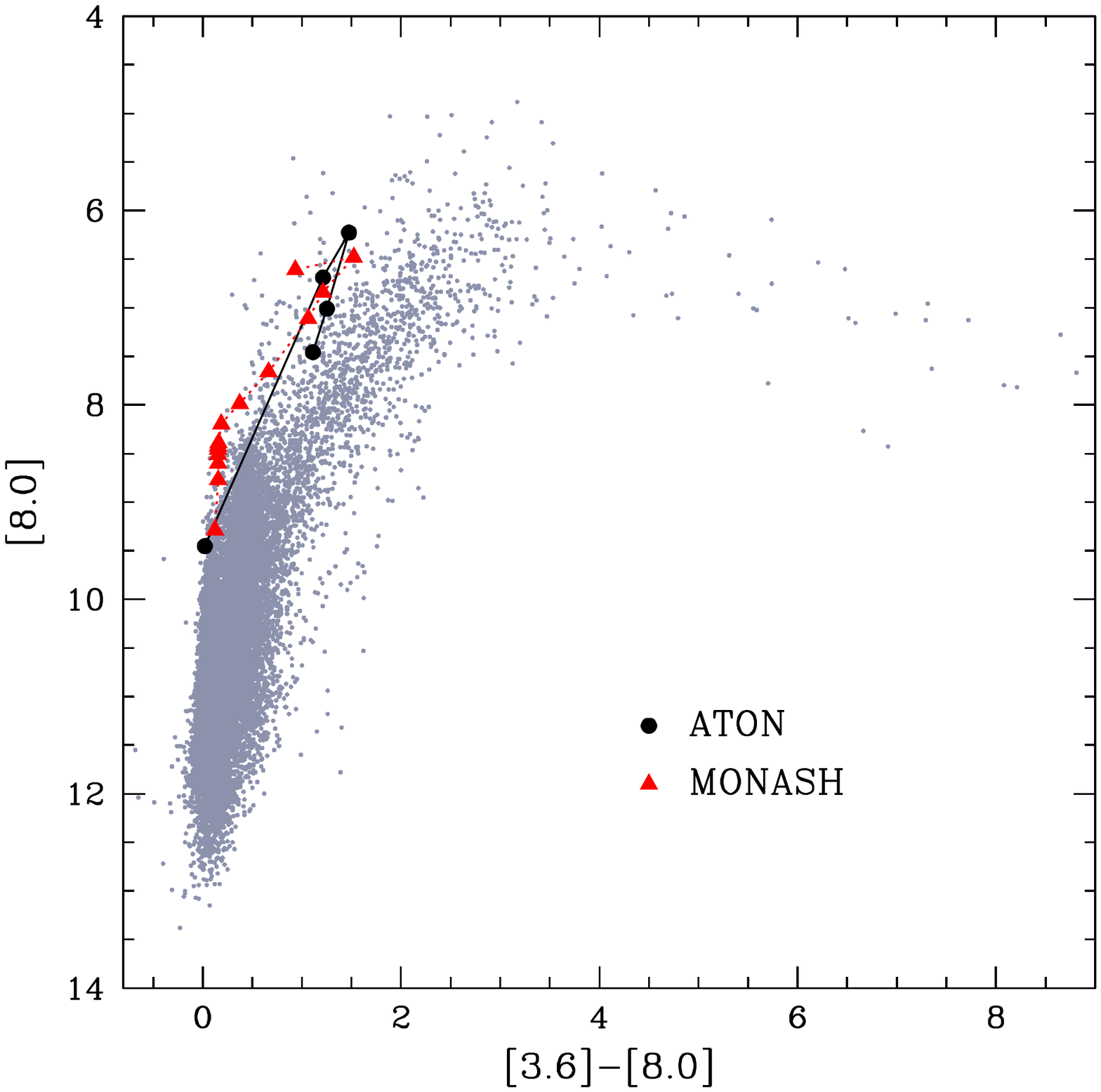}}
\end{minipage}
\vskip-30pt
\caption{The evolution of the ATON and MONASH models of initial mass
$5.5M_{\odot}$ and metallicity $Z=8\times 10^{-3}$ in the colour--colour 
($[3.6]-[4.5]$, $[5.8]-[8.0]$) plane (left panel) and in the colour--magnitude 
($[3.6]-[8.0]$, $[8.0]$) diagram. The points are taken at time intervals of
$3\times 10^4$yr. Grey dots indicate the sample of AGBs in the LMC by 
\citet{riebel12}. The two boxes enclose the regions of the colour--colour plane 
where obscured carbon stars (OCS) and oxygen--rich stars (HBBS) are expected
to evolve, based on the present analysis and on the works by \citet{flavia14a, flavia15}.
}
\label{fccd}
\end{figure*}

\section{Evolutionary properties and Spitzer colours of AGB stars}
\label{spitzer}
In a recent investigation, aimed at interpreting the Spitzer observations of
obscured AGB stars in the LMC, \citet{flavia15} made a characterization of the 
AGB population in terms of age, initial mass and dust properties. The main
result of this study was that carbon and 
oxygen--rich stars with large infrared emission populate two distinct regions in the 
colour--colour ($[3.6]-[4.5]$, $[5.8]-[8.0]$) plane, enclosed within the two boxes shown
in the left panel of Fig. \ref{fccd}; the two groups of stars were defined, respectively, 
OCS (obscured carbon stars) and HBBS (stars experiencing Hot Bottom Burning). 
OCS, with solid carbon and silicon carbide (SiC) particles in their wind, evolve 
along a diagonal band, extending from $[3.6]-[4.5] \sim 0.2$ to $[3.6]-[4.5] \sim 3$.
HBBS, surrounded by alumina and silicates grains, populate the zone on the upper 
side of the diagram, centred at ($[3.6]-[4.5]$, $[5.8]-[8.0]$) $\sim (0.3,0.7)$.

A less evident separation between the two groups of stars is present in the
colour--magnitude ($[3.6]-[8.0]$, $[8.0]$) plane: in this case OCS evolve along a
diagonal band, extending to $[3.6]-[8.0] \sim 6$, whereas HBBS define a more
vertical sequence, with colours $[3.6]-[8.0] < 2$ \citep[see Fig. 11 and Fig. 12 
in][]{flavia15}.

The interpretation by \citet{flavia15} is largely
consistent with samples of spectroscopically confirmed (e.g.,
with Spitzer spectra) AGB stars. However, different results
and/or interpretations may be found among different authors
in the literature. In particular, \citet{riebel12}, based
on best-fit GRAMS models, suggest that $\sim 25\%$ of stars in the HBBS
region are carbon stars. Turning to OCS, interpreted by Dell'Agli et al.
(2015) as a sample entirely composed by carbon stars, preliminary
analysis based on red optical spectra (Boyer et al. 2015, in
preparation) may indicate that a few O-rich sources belong
to the OCS group. These differences stress the need either
for an improvement in the description of dust formation in
the winds of AGBs, and/or a refinement of the classifcation
of AGBs based on red optical spectra.

Understanding the reasons for these discrepancies is far beyond the scope of this work.
However, in the following we will focus on stars in the HBBS region confirmed to be
oxygen--rich.

The ATON model is a typical example of a massive AGB star, belonging to the HBBS sample. 
The comparison between the tracks of the ATON and MONASH models, shown in
Fig. \ref{fccd}, shows a remarkable similarity: during
the phases with the largest infrared emission (corresponding to the shaded regions in
Fig. \ref{fhbb}) the two tracks are practically overimposed in the colour--colour plane, 
a consequence of the similarities in the dust properties. 
The only difference is in the very final AGB phases, when the MONASH
model reaches a C/O ratio greater than one and evolves to the OCS sequence in the 
colour--colour plane, whereas the track of the ATON case stays in the HBBS region.

This is a welcome result, adding more robustness to the conclusion by
\citet{flavia15}. It shows that massive AGB stars, during the phase of highest mass loss
rate, evolve to the HBBS region and that this result is independent of the details of AGB modelling. 
\citet{flavia15} suggested that only stars experiencing strong HBB would populate this region of
the colour--colour plane, whereas here we reach a more general conclusion: all massive 
AGB stars of the metal--rich component in the LMC will evolve into this zone of the diagram.

We return to the main goal of this investigation, i.e., understanding whether the large 
sample of AGB stars in the LMC can be used as a laboratory to test the evolution models of
massive AGB stars and to constrain the strength of the Hot Bottom Burning experienced.
The results found here rule out the possibility that this task can be accomplished on the
basis of pure photometric arguments, because the relevant models evolve to very similar
colours, independently of the details of convection modelling. As shown in the right panel
of Fig. \ref{fccd}, in the colour--magnitude diagram the ATON model reaches lower [8.0] 
magnitudes ($[8.0]\sim 5.7$) compared to the MONASH case ($[8.0]\sim 6.2$); however, the 
statistics in that region of the diagram does not allow to use this information as a valuable 
discriminator between the two descriptions.

The number counts would also be of little help here: despite the overall AGB
phase of the MONASH model is longer, the duration of the phase with 
the largest infrared emission are similar in the two cases (compare the
horizontal extension of the shaded regions in Fig. \ref{fhbb}), thus preventing any
possibility of discriminating based on the number of stars observed in the HBBS 
region. In the MONASH case we expect a larger population of AGBs in the region 
of the colour--colour plane clustering around $[3.6]-[4.5] \sim 0$, where no--dusty, 
oxygen--rich stars evolve. However, this argument cannot be used to 
discriminate among the models: \citet{flavia15} showed that the vast majority of stars 
in that region (see Fig. 15 in \citet{flavia15}) are the progeny of low--metallicity 
stars of mass $1-2M_{\odot}$ (see Fig. 15 in \citet{flavia15}). The fraction of massive 
AGBs present in that region is below $2\%$, 
so on number counts alone it will be 
difficult to distinguish between the ATON and MONASH prescriptions. A better test of
the models is a spectrscopic analysis of the massive AGB stars, using the stars found
from our model independent results for the colour-colour plane.

The predictions of the two models show up significant differences in the dust production
rate (DPR) expected from oxygen--rich and carbon stars. \citet{schneider14} used ATON
models to derive an overall DPR from AGBs of $\sim 4.5 \times 10^{-5} M_{\odot}/yr$, with 
a contribution from carbon and oxygen--rich stars of, respectively, 
$\sim 4 \times 10^{-5} M_{\odot}/yr$ and $\sim 5 \times 10^{-6} M_{\odot}/yr$. Based on 
the MONASH models we expect a smaller relative contribution from oxygen--rich stars.
This is because massive AGB stars are the dominant contributors to the overall silicate 
DPR and the MONASH model is predicted to produce less silicates compared to the ATON
models and to provide a contribution to the overall carbon dust produced, not present in
the ATON case.

While the ATON result for the DPR
from oxygen--rich AGBs is in reasonable agreement with \citet{riebel12} and \citet{matsuura09},
the studies from \citet{srinivasan09} and \citet{boyer12} point in favour of 
MONASH modelling: however, the DPRs in the LMC are estimated using different assumptions
about grain properties, resulting in factors of at least $2-4$ uncertainties, so it is
difficult to make direct comparisons. A reliable estimate of the DPR from AGB stars 
appears as a future, promising indicator of the evolution properties of massive AGB stars.

Despite the similarities in the photometric properties, the surface chemistry of the
ATON and MONASH model during the phase characterised by the large infrared emission 
(indicated with a shaded region in Fig. \ref{fhbb}) is
considerably different. This can be clearly seen in the right--bottom panel of
Fig. \ref{fhbb}, showing the evolution of the surface C/O ratio. In the ATON case the 
chemistry is entirely dominated by the effects of HBB: the C/O ratio is extremely small,
with $C/O < 0.05$, owing to the destruction of the surface carbon via proton fusion.
Conversely, in the MONASH model, the behaviour of the surface chemistry is affected by
both the effects of Third Dredge Up and HBB: in this case the C/O ratio at the surface
of the star is much larger, with $C/O \sim 0.5$, a factor of 10 higher.

We reach the conclusion that the spectroscopic analysis of the stars populating the
region enclosed within the HBBS box in the colour--colour ($[3.6]-[4.5]$, $[5.8]-[8.0]$) 
plane (left panel of Fig. \ref{fccd}), introduced by \citet{flavia15}, 
will be a powerful indicator of the strength of HBB suffered by massive
AGB stars with a metallicity of $Z=8\times 10^{-3}$. The results obtained here are a robust confirm
that these stars are massive AGB stars, surrounded by a thick layer of silicate dust. If the
HBB experienced by these stars is strong, their surface chemistry will reflect the
equilibria of proton capture nucleosynthesis, with a C/O ratio below 0.05; there is no
way to escape from this conclusion, because the stars populating the HBBS region
are expected to be in the phases of strongest Hot Bottom Burning.
On the contrary, a value of $C/O \sim 0.5$ would point in favour of a less efficient 
convection in the envelope, and of a soft HBB.

\citet{mcsaveney07} presented high--dispersion, near--IR spectra of highly evolved 
AGB stars in the SMC and LMC. For a couple of LMC stars the CNO abundances
derived showed--up the signature of HBB, with nitrogen enhancement and carbon deficiency. 
In particular, they measured the C/O ratios in two LMC-HBB AGBs (HV 2576 and NGC 1866 $\#4$), 
obtaining C/O=0.05 and 0.04 in HV 2576 and NGC 1866 $\#4$, respectively.
However, near-IR (H and K bands), high-resolution (R$ > 20,000$)
spectroscopic observations, extended to the O-rich stars in the HBBS sample,
would be needed to confirm the earlier \citet{mcsaveney07} results and to
definitively fix the strength of the HBB experienced by these stars.

\section{Conclusions}
We use the large sample of AGB stars in the LMC to constrain the evolutionary properties of
massive AGB stars, i.e., stars with initial mass $M > 4M_{\odot}$, that are know to experience
Hot Bottom Burning at the base of their convective mantle. The main goal is to draw
information on the strength of the HBB experienced, via a comparison between theoretical
models and observations.

To this aim, we compare results from two independent research groups involved in AGB studies,
whose models of massive AGB stars are known to differ in the efficiency of the convection
modelling, and consequently in the HBB experienced.

The dust formation process in the winds of these stars is found to be essentially independent
of the details of AGB modelling. During the phases when the stars evolve at the highest
luminosity, a common behaviour of the different models is the formation of alumina dust
and silicates grains. The former species is more stable and form in a more internal region,
with the grain size distribution peaked at $\sim 0.05\mu$m, partly dependent on the AGB 
model used. Silicate particles of $\sim 0.12\mu$m size form in a more external region;
owing to their large extinction coefficients, this dust species favour the acceleration
of the wind, up to velocities of the order of $10-15$ km s$^{-1}$.

Owing to the similarity in the dust composition surrounding the stars, the infrared
colours expected are also practically independent of AGB modelling. 

This result is in agreement with recent investigations and offer the opportunity of
selecting a well defined region in the colour--colour ($[3.6]-[4.5]$, $[5.8]-[8.0]$) plane 
where massive AGBs, experiencing Hot Bottom Burning, evolve.
More important, this finding indicates the possibility of an observational test
of the efficiency of the Hot Bottom Burning experienced by AGBs, at least at the metallicity
typical of young LMC stars. This analysis will be possible because we have shown that
the surface chemical composition of massive AGB stars during the phase with the 
highest infrared emission is extremely sensitive to convection modelling: we suggest that
a near-IR (H and K bands) spectroscopic follow--up of the stars identified as massive 
AGBs based on their position on the aforementioned plane would provide a clear and 
straight indication of the strength of HBB experienced.

\section*{Acknowledgments}
P.V. was supported by PRIN MIUR 2011 "The Chemical and Dynamical Evolution of the Milky Way 
and Local Group Galaxies" (PI: F. Matteucci), prot. 2010LY5N2T. A.I.K. was supported 
through an Australian Research Council Future Fellowship (FT110100475). D.A.G.H. 
acknowledges support provided by the Spanish Ministry of Economy and Competitiveness under 
grant AYA-2011-27754. R.S. acknowledges funding from the European Research Council under 
the European Unionâs Seventh Framework Programme (FP/2007- 2013)/ERC Grant Agreement 
n. 306476.

\end{document}